%% file: main.tex
\documentclass[sigconf]{acmart}
\AtBeginDocument{%
  }

\usepackage{subcaption}
\usepackage{algorithm}
\usepackage{algorithmic}
\usepackage{hyperref}
\usepackage{multirow}

\usepackage{listings}

\lstdefinestyle{cuda}{
  language=C++, 
  basicstyle=\ttfamily\footnotesize, 
  keywordstyle=\color{blue}\bfseries, 
  commentstyle=\color{green!50!black}\itshape, 
  stringstyle=\color{orange}, 
  numbers=left, 
  numberstyle=\tiny\color{gray}, 
  stepnumber=1, 
  numbersep=5pt, 
  backgroundcolor=\color{gray!10}, 
  frame=single, 
  breaklines=true, 
  tabsize=2, 
  morekeywords={m256, m256i, m256_load, m256_store_ps, m256_mul_ps, m256_gather_ps, m256_cast_ps, m256_add_ps, m256_fma_ps, m256_sub_ps, m256_set_ps, m256_div_ps, m256_blend_ps, m256_cmpeq,__shfl_inst, warp_inclusive_scan}, 
  morecomment=[l][\color{magenta}]{//}, 
  morecomment=[s][\color{magenta}]{/*}{*/}, 
  morestring=[b]" 
}

\setcopyright{acmlicensed}
\copyrightyear{2026}
\acmYear{2026}
\setcopyright{cc}
\setcctype{by}
\acmConference[HPDC '26]{The 35th International Symposium on High-Performance Parallel and Distributed Computing}{July 13--16, 2026}{Cleveland, OH, USA}
\acmBooktitle{The 35th International Symposium on High-Performance Parallel and Distributed Computing (HPDC '26), July 13--16, 2026, Cleveland, OH, USA}
\acmDOI{10.1145/3806645.3807593}
\acmISBN{979-8-4007-2640-8/2026/07}

\begin{document}

\title{Endeavor: Efficient PairHMM for Detection of DNA Variants in Genome-Scale Datasets}


\author{Miguel Graça}
\email{miguel.graca@inesc-id.pt}
\affiliation{%
  \institution{INESC-ID, Instituto Superior Técnico}
  \city{Lisbon}
  \country{Portugal}
}

\author{Aleksandar Ilic}
\email{aleksandar.ilic@inesc-id.pt}
\affiliation{
  \institution{INESC-ID, Instituto Superior Técnico}
  \city{Lisbon}
  \country{Portugal}
}


\begin{abstract}
DNA variant calling represents a key operation in bioinformatics pipelines that aims at identifying genetic variants. Given an evidenced explosion in genomic data availability, there is an urgent need for a high-performant, portable and efficient solution for variant calling, which can further improve our understanding of genomic structure and genetic basis for complex diseases. In its most common formulation, the Pair Hidden Markov Model (PairHMM) algorithm for variant calling stands as the main bottleneck in the pipeline, accounting for up to 70\% of the execution time in large-scale genomic datasets. The state-of-the-art approaches for accelerating PairHMM in CPUs and GPUs do not scale to long DNA sequences and only explore very limited anti-diagonal data parallelism, which yields poor performance. In this work, Endeavor is proposed as a new parallelization strategy for PairHMM that redefines its traditional formulation to explore row-level fine-grained parallelism without loss in solution accuracy. Based on this, a novel and portable SIMD-based approach is derived for efficient and high-performance processing of short and long sequences in CPUs and GPUs, leveraging novel levels of parallelism and synchronization to achieve high throughput in sequences up to 100k basepairs for the first time. Evaluation on Intel and AMD CPUs shows that Endeavor outperforms GKL up to 2.14x in peak throughput and GATK HaplotypeCaller by at least 2x in real-world datasets, while NVIDIA and AMD GPUs achieve up to 2.05x speedups in genome-scale datasets when compared to state-of-the-art GPU-based methods.

\end{abstract}

\ccsdesc[500]{Applied computing~Computational biology}
\ccsdesc[500]{Computer systems organization~Parallel architectures}
\ccsdesc[500]{Computing methodologies~Parallel algorithms}

\keywords{Variant Calling, Parallel Computing, Bioinformatics}


\maketitle

\section{Introduction}
\label{sec:Introduction}
\input{tex/Introduction}

\section{Background and Related Work}
\label{sec:Background}
\input{tex/Background}

\section{Proposed Hardware Acceleration for Variant Calling} 
\label{sec:Methods}
\input{tex/Methods}

\section{Experimental Results} \label{exp}
\label{sec:Results}

\input{tex/Results}

\section{Conclusions} 
\label{sec:Conclusion}
\input{tex/Conclusion.tex}

\begin{acks}
\label{sec:Acks}
\input{tex/Acks}
\end{acks}

\bibliographystyle{ACM-Reference-Format}
\bibliography{acmart}

\end{document}

%% file: tex/Introduction.tex
The Human Genome Project \cite{gibbs2020human}, set to sequence more than 90\% of the human genome, represented a milestone in genomics that took 13 years to achieve (completed in 2003). During the following 20 years, Next-Generation Sequencing (NGS) technologies \cite{hu2021next} have paved the way to sequence genomes faster and cheaper. This rapid technological advance culminated in the development of more ambitious genetic projects, such as the 1000 Genomes Project \cite{10002015global} and the Genome In A Bottle (GIAB) consortium \cite{xiao2014giab}, as well as an unprecedented explosion of the available genomic data. 

The widespread availability of such data led to the emergence of a novel bioinformatics pipeline, known as variant calling \cite{olson2023variant}, which analyses the existence of genetic dissimilarities between the patient's genome and a reference genome. Variant calling is of the utmost importance to identify genetic differences and understand their correlation with complex diseases \cite{koboldt2020best}, which has a significant impact in precision medicine \cite{ashley2016towards}, pharmacogenomics \cite{giannopoulou2019integrating}, and evolutionary biology \cite{jonsson2017whole}. As a consequence, specific toolkits have been developed to process genetic data for variant calling, with the Genome Analysis Toolkit (GATK) HaplotypeCaller \cite{depristo2011framework} being one of the most widely used in genomic analysis.

The typical workflow of variant calling in GATK follows three steps. Given a dataset with read sequences from a patient, the first step is to generate a set of candidate haplotypes by assembling the reads \cite{ren2017gpuaccel}. The second step is to calculate the probability that a read was generated by a given haplotype. To this end, the Pair Hidden Markov Model (PairHMM) algorithm \cite{yoon2009hidden} aligns the read sequence data and the candidate haplotypes, calculating the alignment likelihoods for each candidate and selecting the haplotypes with the highest likelihood. Finally, a Bayesian model is applied to the computed likelihoods to determine the most probable genotype at a given read (if it is equal to a reference genome or a variant). 

Of these three steps, the PairHMM algorithm is the main bottleneck, responsible for more than 70\% of the execution time in genome-scale datasets \cite{li2021improved}, and therefore is the focus of hardware acceleration efforts. However, the PairHMM algorithm is not trivial to accelerate and the main reasons for this are two-fold. First, PairHMM scales quadratically in time and memory with the sequence size (as it is based on computing matrix cells), making it prohibitive for sequences generated with long-read sequencing technologies. Although memory requirements can be reduced to scale linearly with sequence length \cite{sampietro2018fpga}, the same cannot be done for time requirements. Second, the algorithm uses dynamic programming \cite{eddy2004dynamic}, which hinders the opportunities to parallelize the algorithm due to dependencies in the data flow.

Nevertheless, various works in the literature attempted to parallelize PairHMM by leveraging the anti-diagonal independence and implementing solutions on FPGAs \cite{sampietro2018fpga, huang2017hardware, banerjee2017accelerating} and GPUs \cite{li2021improved, branchini2021methodology, ren2018efficient}. However, the current solutions suffer from three major drawbacks. First, CPUs are still the de-facto standard in various bioinformatics applications and tools \cite{buchfink2021sensitive, zhang2026faster,minh2020iq,camacho2023elasticblast}. While highly-optimized PairHMM CPU-based implementations \cite{depristo2011framework} do not offer the performance of massively parallel GPUs, they can evaluate long-read datasets (where the sequence length can exceed $10^4$ to $10^5$ basepairs \cite{jain2018nanopore}). Second, the current GPU-based solutions are not scalable with the sequence length, i.e., they cannot deal with long-read datasets, and are not portable. The existent GPU approaches target only NVIDIA GPUs which, given the current heterogeneous computational landscape, lack the necessary portability to target different hardware devices (e.g., the top 5 fastest supercomputers in the world \cite{top500} include AMD MI300A GPUs, which cannot be targeted for PairHMM acceleration with the existing solutions from the literature). Finally, the anti-diagonal parallelization strategy is not efficient on both architectures, due to the variable amount of computations that the algorithm performs as it traverses the matrices' anti-diagonals. Therefore, one of the main research challenges nowadays is how to further improve our understanding of genomes \cite{nurk2022complete} by efficiently processing data from both short- and long-read sequencing technologies and leveraging different hardware architectures to achieve portability.

This is a particular gap that this paper intends to close by proposing Endeavor (\textbf{E}fficie\textbf{N}t PairHMM for \textbf{DE}tection of DN\textbf{A} \textbf{VAR}iants), a novel parallelization approach for the PairHMM algorithm on CPUs and GPUs that fully leverages the hardware resources and scales to long DNA sequences. The main contributions of this paper are as follows:

\begin{enumerate}
    \item Redefinition of the processing steps of the PairHMM algorithm to expose previously unexplored levels of parallelism with no loss in solution accuracy;
    \item An efficient, portable, and fine-grained parallel approach to achieve high performance on CPUs and GPUs;
    \item Scalability to diverse genomic datasets with short- and long-read sequences.
\end{enumerate}

Contribution (1) pertains to the novel parallelization strategy that is herein proposed to accelerate PairHMM, which leverages row-level parallelism of the matrices to fully exploit performance at the hardware level (with no loss in solution precision when compared to the anti-diagonal approach).

Contribution (2) is concerned with the efficient mapping of the proposed approach to extract data parallelism. To benefit from the row-level parallelism on the CPU, the proposed solution focuses on multi-threading and SIMD intrinsics to process read-haplotype pairs at the level of each SIMD lane. On the GPU, Endeavor focuses on shared memory and cross-thread intrinsics to efficiently map the PairHMM computations, which are implemented in CUDA and HIP to target GPUs from major vendors (NVIDIA and AMD). To achieve unprecedented performance for long reads, novel features from recent GPU architectures, namely inter-thread-block operations, are also efficiently leveraged.

Contribution (3) focuses on Endeavor's portability and scalability across different devices. Compared to the Intel Genomics Kernel Library (GKL) \cite{foley2017accelerate}, Endeavor achieves up to 2.14x improvements in peak throughput in different CPU architectures, while evaluation on genome-scale datasets shows that Endeavor outperforms a highly optimized (OpenMP + AVX512) GATK HaplotypeCaller's PairHMM implementation on CPUs by at least 2x in GIAB datasets. Compared with a state-of-the-art PairHMM GPU-based method (referred herein as gpuPairHMM \cite{schmidt2024gpupairhmm}) that only employs single floating-point precision on vendor-specific GPUs, Endeavor achieves speedups up to 2.05x on large-scale datasets, while providing support for both single- and double-precision, portability across different GPU architectures, and processing sequences up to 100k basepairs, representing an increase by 1 order of magnitude compared to the literature and demonstrating its flexibility to process short- and long-read datasets.

%% file: tex/Background.tex
Variant calling aims to identify genetic variants by comparing a DNA sequence (a string of nucleotides) from a patient (herein referred to as a read sequence) and a reference DNA sequence (herein referred to as a haplotype sequence) \cite{koboldt2020best}. To achieve this, the PairHMM algorithm computes all possible alignments between the read and the haplotype. An alignment describes the relationship between two sequences, where characters on both strings can be aligned to each other or to an empty space (herein referred to as gap) due to inserted or deleted characters on one of the strings. 

As a probabilistic model, PairHMM attributes a probability for each alignment \cite{durbin1998biological}. The algorithm’s output is the aggregated probability of all alignments, which describes the likelihood that the read sequence is derived from the haplotype sequence. To perform this calculation, PairHMM genomic datasets provide the read-haplotype pairs, and four additional strings (of the same length as the read). These strings provide the read's quality score (the probability that a given character in the read is erroneous), the insertion and deletion probabilities (the likelihood of an inserted or deleted character in the read when compared to the haplotype), and the gap probabilities in the sequences. 

\begin{figure}[t]
\centering
\includegraphics[width=\linewidth]{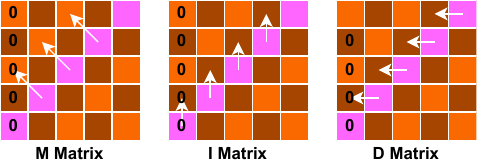}
\caption{Antidiagonal Dependencies of PairHMM.}
\label{fig:antidiagonal}
\end{figure}

Given a read and haplotype sequences, of length $l$ and $k$, respectively, the PairHMM algorithm operates on three matrices ($M$, $I$, and $D$) and is initialized as follows 
\begin{equation}
    M_{i,0} = I_{i,0} = D_{i,0} = M_{0,j} = I_{0,j} = 0, D_{0,j} = 2^{1020}/k,
    \label{eq:init}
\end{equation}
where the first row and column of $M$ and $I$ are initialized to 0 and $D$ is initialized to a large constant to prevent underflow. The recurrence equations for the matrices are given by
\begin{equation}
    \begin{aligned}
    M_{ij} & = P(r_i|h_j)(T_{MM_{i}}M_{i-1,j-1} + T_{IM}I_{i-1,j-1} + T_{DM}D_{i-1,j-1}),\\
    I_{ij} & = T_{MI_{i}}M_{i-1,j} + T_{II}I_{i-1,j},\\
    D_{ij} & = T_{MD_{i}}M_{i,j-1} + T_{DD}D_{i,j-1},
    \end{aligned}
    \label{eq:recurrence}
\end{equation}
where a read position is fixed ($i$), while all haplotype positions are iterated ($j$). In detail, $M_{ij}$ is the overall probability of sub-sequences $r_1,...,r_i$ and $h_1,...,h_j$ when $r_i$ aligns to $h_j$, while $I_{ij}$ and $D_{ij}$ are the overall probability when $r_i$ or $h_j$ aligns to a gap, respectively. $T_{MM_{i}}, T_{MI_{i}}$, and $T_{MD_{i}}$ represent transition probabilities that vary with the read position, $r_i$. Conversely, the probabilities $T_{DD}, T_{II}, T_{DM}$, and $T_{IM}$ are set to be constant, with $T_{DD} = T_{II} = 0.1$ and $T_{DM} = T_{IM} = 0.9$. These values come from the input insertion, deletion, and gap probabilities. Lastly, the term $P(r_i|h_j)$ represents the conditional probability of the read at position $i$, given the haplotype at position $j$, calculated as
\begin{equation}
  P(r_i|h_j) =
    \begin{cases}
      10^{\frac{-(Q_i - 33)}{10}}/3 & \text{if $r_i \neq h_j$}\\
      1 - 10^{\frac{-(Q_i - 33)}{10}} & \text{if $r_i = h_j$}\\
    \end{cases}  
    \label{eq:quality}
\end{equation}
where $Q_i$ is a base quality score for the read at position $i$. The final result is a likelihood, $L$, which is the cumulative probability of all sequence alignments, calculated as
\begin{equation}
    L = \sum_j M_{l,j} + I_{l, j},\\
    \label{eq:likelihood}
\end{equation}
which depends only on the last rows of $M$ and $I$. Figure \ref{fig:antidiagonal} displays how the data dependencies of PairHMM imply the parallel processing opportunities only in antidiagonals. When calculating an element of $M$, $I$, or $D$, the dependencies are on the diagonal, column, and row elements, respectively. Therefore, an anti-diagonal can be completely parallelized, using previous anti-diagonals. 

To accelerate the PairHMM algorithm, previous works in the literature have leveraged this property for CPU, GPU, and FPGA's parallelization. One of the most well known implementations in CPUs is included in the Intel Genomics Kernel Library (GKL) \cite{foley2017accelerate}, which uses AVX and AVX-512 \cite{snytsar2023pairhmm} instructions for SIMD parallelization of the matrices' anti-diagonals, as well as OpenMP to extract the most performance out of multicore processors. While this implementation is used in GATK HaplotypeCaller \cite{depristo2011framework}, the anti-diagonal strategy has two major drawbacks: the uncoalesced memory access pattern on the matrices and the variable computations (the first and last anti-diagonals, in the top-left and bottom-right section of the matrices, have significantly fewer elements to process than the anti-diagonals in the center). Specialized FPGA-based hardware designs have also been proposed to explore anti-diagonal dependencies \cite{wertenbroek2019acceleration, peltenburg2016maximizing, chen2019banded} and reduce the memory footprint \cite{sampietro2018fpga, huang2017hardware, banerjee2017accelerating}. Although these approaches typically provide better energy-efficiency, their limited computational resources are prohibitive for sequences with long lengths \cite{robinson2021hardware}, which hinders their scalability. 

Due to their massively parallel architecture, GPUs are a prominent solution to accelerate PairHMM that also rely on the anti-diagonal independence \cite{ren2018efficient, branchini2021methodology}, although this approach is not ideal, given the variance of values to calculate on each anti-diagonal and the memory access pattern. In \cite{li2021improved}, a parallelization strategy is proposed by conceptualizing a processing unit to evaluate anti-diagonal cells with an adjustable number of threads. By setting the number of threads to a small number (e.g., 4), the first and last anti-diagonal do not have many idle GPU threads, improving throughput. Additionally, batch parallelization is also proposed to process many sequence pairs simultaneously. However, the implementation assumes that the read lengths do not exceed 300, since it was tailored for Illumina \cite{luo2012direct} datasets. Therefore, it does not scale for long-read datasets, which have sequence lengths $> 10^4$. Frameworks that are tailored for dynamic programming, such as CUDA Dynamic Parallelism (CDP) \cite{wang2014characterization, adinetz2014adaptive}, have also been explored \cite{liu2023genomics}. CDP allows a CUDA kernel (parent) to launch new kernels (children) without returning control to the host, allowing for nested parallelism. This programming model was used to accelerate the PairHMM algorithm \cite{liu2023genomics}, but the kernel launch overhead, as well as the hardware limits on the number of concurrent kernel launches and the depth of nested kernels \cite{carneiro2018gpu}, significantly impact the scalability of CDP for long-read datasets. In the current literature, a novel approach herein referred as gpuPairHMM \cite{schmidt2024gpupairhmm} is the fastest PairHMM implementation, leveraging the anti-diagonal independence and processing read-haplotype pairs at the level of a warp / subwarp by having individual threads calculate multiple elements of $M$, $I$, and $D$ to achieve unprecedented throughput for PairHMM.

While these solutions achieve good results on short-read genomic datasets, the current solutions are vendor-specific and based on a non-ideal parallelism pattern. Therefore, a breakthrough in performance is expected if a PairHMM approach extracts more parallelism from the algorithm while keeping a high throughput and scaling to long reads to process any real genomic dataset in different devices. To this end, Endeavor proposes a row-wise parallelism strategy as a possible solution, which has been applied to the sequence alignment problem \cite{khajeh2010acceleration}, but has never been studied for PairHMM. An ideal row-wise definition would do one pass per row, but the existing mathematical proofs for sequence alignment require two passes per row \cite{aluru2003parallel} and the key operations are different from PairHMM (maximums instead of sums and products). Although both algorithms share similarities, PairHMM brings its own unique parallelization challenges, which Endeavor addresses in the CPU and GPU methods proposed in this work (Section \ref{sec:Methods}) to deliver high-performance solutions for different hardware architectures to overcome the limitations of the state-of-the-art.

%% file: tex/Methods.tex
In this section, the proposed CPU and GPU-based approaches to accelerate PairHMM is described. In particular, a mathematical derivation starting from the algorithm's analytical formulation is provided, as it is an essential step to understand how we extract more fine-grained data parallelism. It is worth noting that the proposed algorithm is a novel approach to PairHMM parallel processing, which overcomes the limitations of existing state-of-the-art methods that only exploit parallelism on anti-diagonals (see Section \ref{sec:Background}). Afterwards, we elaborate on how the proposed approach is employed to define the proposed methods to process short-read datasets (sequences with a range of 250-800 basepairs), as well as long reads (sequences with $> 10^4-10^5$ basepairs).

\subsection{Algorithm Design}

\begin{figure}[t]
\centering
\includegraphics[width=0.9\linewidth]{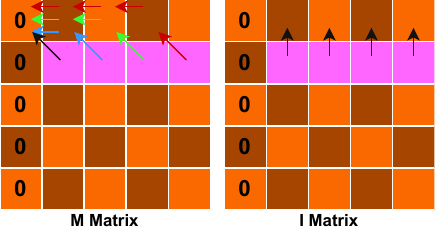}
\caption{Rowwise Dependencies of Endeavor.}
\label{fig:row}
\end{figure}

As mentioned in Section \ref{sec:Background} (see Equation \ref{eq:recurrence}), there are three main expressions that characterize the PairHMM algorithm and the dependencies between the $M$, $I$, and $D$ matrices. One aspect worth noticing about PairHMM is that the final result (see \ref{eq:likelihood}) does not require the $D$ matrix and is only dependent on $M$ and $I$. However, $D$ is still necessary for the intermediate calculation of the elements of $M$ (see $D_{i-1,j-1}$ in Equation \ref{eq:recurrence}). Based on this observation, we start by expressing $D_{i-1,j-1}$ as a function of $M$ and $I$ (as stated in Equation \ref{eq:recurrence}) such that 
\begin{equation}
    D_{i-1,j-1} = T_{MD_{i-1}}M_{i-1,j-2} + T_{DD}D_{i-1,j-2}.\\
\end{equation}
By relying on the same expression, the first four terms (i.e., $j \in \{1,...,4\}$) of the $i-1$ row of $D$ can be expressed as
\allowdisplaybreaks
\begin{align*}
D_{i-1,1} & = T_{MD_{i-1}}M_{i-1,0} + T_{DD}D_{i-1,0} = 0\\
D_{i-1,2} & = T_{MD_{i-1}}M_{i-1,1} + T_{DD}D_{i-1,1} = T_{MD_{i-1}}M_{i-1,1}\\
D_{i-1,3} & = T_{MD_{i-1}}M_{i-1,2} + T_{DD}D_{i-1,2}\\ 
& = T_{MD_{i-1}}M_{i-1,2} + T_{DD} T_{MD_{i-1}}M_{i-1,1}\\
& = T_{MD_{i-1}}(M_{i-1,2} + T_{DD}M_{i-1,1})\\
& = T_{MD_{i-1}}\sum_{b=1}^{2}M_{i-1,b}T_{DD}^{2 - b}\\
D_{i-1,4} & = T_{MD_{i-1}}M_{i-1,3} + T_{DD}D_{i-1,3}\\ 
& = T_{MD_{i-1}}(M_{i-1,3} + T_{DD}M_{i-1,2} + T_{DD}^2 M_{i-1,1})\\
& = T_{MD_{i-1}}\sum_{b=1}^{3}M_{i-1,b}T_{DD}^{3 - b},\\
\label{eq:unroll}
\end{align*}
where $D_{i-1,1} = 0$ as $M_{i-1,0} = D_{i-1,0} = 0$ (see Equation \ref{eq:init}). If the unrolling procedure is applied to each new term of $D$, one obtains a general sum of values in $M_{i-1,b}$ that can be written as follows 
\begin{equation}
    \begin{aligned}
    D_{i-1,j-1} & = T_{MD_{i-1}}(M_{i-1,j-2} + T_{DD}M_{i-1,j-3} + ... + T_{DD}^{j-3}M_{i-1,1}) \\
    & = T_{MD_{i-1}}\sum_{b=1}^{j-2}M_{i-1,b}T_{DD}^{j - 2 - b}.\\
    \end{aligned}
    \label{eq:dloop}
\end{equation} 
As can be observed, Equation \ref{eq:dloop} shows that $D_{i-1,j-1}$ depends only on previous values of $M_{i-1, b}$. Therefore, both $M_{ij}$ and $I_{ij}$ (see Equation \ref{eq:recurrence}) depend only on values from the previous row, $M_{i-1,b}$ and $I_{i-1,b}$. As such, all the elements in a row can be calculated in parallel. In this case, the equations for PairHMM can be expressed as follows
\begin{equation}
    \begin{aligned}
    M_{ij} & = P(r_i|h_j)(T_{MM_{i}}M_{i-1,j-1} + T_{IM}I_{i-1,j-1}\\
           & + T_{DM}T_{MD_{i-1}}\sum_{b=1}^{j-2}M_{i-1,b}T_{DD}^{j - 2 - b})\\
    I_{ij} & = T_{MI_{i}}M_{i-1,j} + T_{II}I_{i-1,j},\\
    \end{aligned}
    \label{eq:dloop2}
\end{equation}
substituting Equation \ref{eq:recurrence}. Defining PairHMM in this fashion provides two major benefits, as is shown in Figure \ref{fig:row}. First, $D$ is no longer necessary, which reduces the storage requirements for PairHMM. Second, removing $D$ eliminates one of the dependencies, which improves parallelism and performance. For a read of length $l$ and a haplotype of length $k$, the anti-diagonal implementation calculates $M$, $I$, and $D$, requiring $\mathcal{O}(l + k - 1)$ passes (since there are $l + k - 1$ antidiagonals). In comparison, the proposed approach calculates only $M$ and $I$ (which are the necessary matrices for the likelihood), requiring only $\mathcal{O}(l)$ passes to do all calculations. Furthermore, the proposed algorithm outputs the same results as GATK HaplotypeCaller \cite{depristo2011framework}, without any loss in accuracy.

\subsection{Endeavor Approach Overview} \label{sec:overview}

\begin{figure}[t]
\centering
\includegraphics[width=\linewidth]{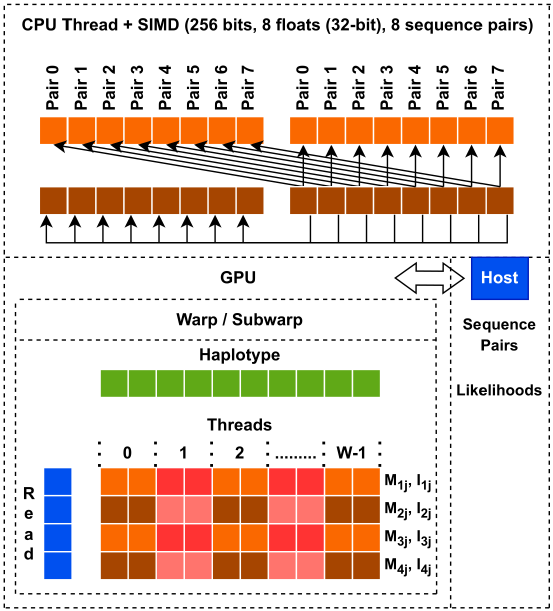}
\caption{Multithreading+SIMD-Based (CPU) and Warp-Based (GPU) Implementations of PairHMM.}
\label{fig:threadblock}
\end{figure}

To map the redefined PairHMM algorithm to multicore CPUs and highly parallel GPU hardware, a high level overview of Endeavor is provided herein and described in Figure \ref{fig:threadblock}. At the CPU level, read-haplotype pairs are read from a dataset and processed at the level of a single SIMD lane (e.g., in Figure \ref{fig:threadblock}, using double precision (64-bit) and a SIMD width of 256 bits allows a single thread to process 4 (256/64) read-haplotypes in parallel). As read-haplotype pairs are independent, the use of multiple threads, each one operating on SIMD lanes, allows to maximize the algorithm's throughput.

At the GPU level, the host reads read-haplotype pairs from a dataset, which are sent to the GPU. After performing the PairHMM computations, the GPU sends the likelihoods back to the host. Endeavor's approach aims at minimizing synchronization in GPUs by processing the sequences at the level of a warp of size $W$ (where $W = 32$ if the GPU is from NVIDIA or Intel and $W = 64$ if the GPU is from AMD to match the warp size on each architecture). All threads iterate over the read's characters to calculate the rows of $M$ and $I$, while each individual thread reads $N$ characters from the haplotype and calculates $N$ elements of a $M$ and $I$ row (e.g., in Figure \ref{fig:threadblock}, each thread is responsible for calculating 2 elements of a row). After calculating the last row, the values of $M$ and $I$ are accumulated across all threads to obtain the final likelihood, which is written to global memory and sent back to the host.

\begin{lstlisting} [float,style=cuda, caption={CPU kernel for AVX-level PairHMM (FP32).}, label=code:cpupairhmm]
template <unsigned int K> 
void avx_fp32(int* hap, int* read, int* _q, int* _i, int* _d, int seqsize, float* sum, float* q32)
{
  #pragma omp parallel
  { 
    int tid = omp_get_thread_num();
    int nthr = omp_get_num_threads();

    for(int seq = tid; seq < seqsize/8; seq += nthr){
      m256 Mp[K], Ip[K], _hap[K];
      m256 M, y2, y3, one, thr, d0, tmm, pr, qual, tmi, mask;
      m256 one0 = m256_set_ps(0.1f), nine = m256_set_ps(0.9f), finalsum = m256_set_ps(0.0f);
      m256i rs, q0, in, d1;

      rs = m256_load(read);
      q0 = m256_load(q);
      d1 = m256_load(d);
      qual = m256_gather_ps(q32, q0, 4);
      thr = m256_mul_ps(m256_set_ps(1/3),qual);
      one = m256_sub_ps(m256_set_ps(1),qual);
            
        for(int h = 0; h < K; h++){
          _hap[h] = m256_load(hap + h*8)
          mask = m256_cast_ps(m256_cmpeq(rs, _hap[h]));
          pr = m256_blend_ps(thr, one, mask);
          Mp[h] = m256_mul_ps(pr, m256_mul_ps(nine, m256_div_ps(m256_set_ps(ldexpf(1.f, 120.f)),m256_set_ps(N))));
          Ip[h] = m256_set_ps(0);
        }  

        for(int k = 1; k < K; k++){
          rs = m256_load(read + k*8);
          q0 = m256_load(q + k*8);
          in = m256_load(i + k*8);
          d0 = m256_mul_ps(nine,m256_gather_ps(q32, d1, 4));
          d1 = m256_load(d + k*8);

          mask = m256_cast_ps(m256_cmpeq(rs, _hap[0]));
          qual = m256_gather_ps(q32, q0, 4);
          tmi = m256_gather_ps(q32, in, 4);
          thr = m256_mul_ps(m256_set_ps(1/3),qual);
          one = m256_sub_ps(m256_set_ps(1),qual);
          pr = m256_blend_ps(thr, one, mask);
          tmm = m256_sub_ps(m256_sub_ps(m256_set_ps(1), m256_gather_ps(q32, d1, 4)), tmi);
          y3 = M = m256_set_ps(0);
                
          for(int h = 1; h < K; h++){
            mask = m256_cast_ps(m256_cmpeq(rs, _hap[h]));
            pr = m256_blend_ps(thr, one, mask);
            y2 = m256_mul_ps(pr, m256_add_ps(m256_add_ps(m256_mul_ps(Mp[h-1],tmm), m256_mul_ps(nine,Ip[h-1])), m256_mul_ps(d0,y3)));
            y3 = m256_fma_ps(y3,one0,Mp[h-1]);    
            Ipr[h-1] = m256_fma_ps(Mp[h-1],tmi, m256_mul_ps(Ip[h-1],one0)); 
            Mp[h-1] = M;
            M = y2;      
          }    

          Ip[N-1] = m256_fma_ps(Mp[N-1],tmi, m256_mul_ps(Ip[N-1],one0)); Mp[N-1] = M;
        }
        
        for(int h = 0; h < K; h++) finalsum = m256_add_ps(finalsum, m256_add_ps(Mp[h], Ip[h]));

        m256_store_ps((sum + seq), finalsum);
    }
  }
}
\end{lstlisting}

To achieve high throughput, Endeavor focuses on several different aspects. At the GPU level, read-haplotype sequence pairs can also be processed at the level of a sub-warp of size $W/P$. As an example, if a single thread computes $N$ elements of a row of $M$ and $I$ and haplotypes of size $2N$ are to be analyzed, then only 2 threads are necessary for the computations, which allows 16 read-haplotype pairs to be processed if $W = 32$ and $P = 16$ or 32 read-haplotype pairs if $W = 64$ and $P = 32$. It is worth noting that all threads within a warp will still adhere to SIMD processing with no divergency (i.e., performing the same operations on different data). To optimize computations for different sequence lengths, read-haplotype pairs are batched according to their lengths and run on dedicated kernels according to the value of $P$, which defines the warp/subwarp size ($W/P$). To maximize performance, the value of $N$ must be carefully selected to ensure that each thread has enough computations to perform and to avoid register spilling to global memory, which would decrease throughput. In Section \ref{sec:kernel}, an analysis of the impact of $N$ in different hardware architectures is performed. 

At the CPU level, the SIMD width impacts the read-haplotype pairs that each thread can process in parallel. Endeavor explores the impact in performance of two different SIMD extensions: AVX (256 bits) and AVX-512 (512 bits). Finally, for both devices, as the computations for PairHMM revolve around probabilities, precision impacts the algorithm's accuracy and maximum throughput. To this end, Endeavor implements PairHMM in double and single precision. In Section \ref{sec:scale}, Endeavor's peak throughput is analyzed in these two precisions for various CPU and GPU architectures.

\subsection{CPU SIMD Implementation}

Listing \ref{code:cpupairhmm} provides an example of the proposed CPU implementation to process read-haplotype pairs of size $K$ in single precision, using AVX intrinsics, as described in Figure \ref{fig:threadblock} (note that, since AVX performs computations with 256 bits and single precision values (32-bit) are used, this amounts to processing 8 (256/32) sequence pairs per thread). The main inputs are the read and haplotype sequences, the quality ($\_q$), insertion ($\_i$), and deletion ($\_d$) scores (described as characters, which are associated to powers of 10, as demonstrated in Equation \ref{eq:quality}), an array with powers of 10 ($q32$), and the total number of sequence pairs to process ($seqsize$). The output is an array of sums (each position will hold the sum of the last rows of $M$ and $I$ for a given read-haplotype pair). For this kernel, it is assumed that all the evaluated sequences have length $K$.

\begin{figure}[t]
\centering
\includegraphics{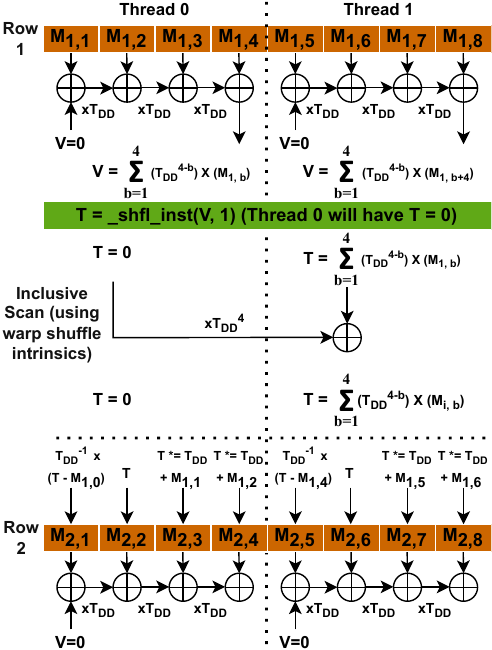}
\caption{Endeavor's GPU Pipeline for $M$ (for this example, $N = 4$, with 2 threads per read-haplotype pair).}
\label{fig:general}
\end{figure}

In line 4, the kernel spawns multiple threads to perform parallel work. Lines 6 to 14 initialize the necessary variables with which each thread will work, namely the $M$ and $I$ rows ($Mp$, $Ip$) and haplotype characters ($\_hap$) in line 10. Note that $Mp$ and $Ip$ are defined as $m256$ arrays of size $K$ where, for example, the first 32 bits in each position define the $M$ and $I$ rows for the first read-haplotype pair to process. The first read and quality characters are read in lines 15 to 17. In line 18, a gather operation ($m256\_gather\_ps$) loads from memory the powers of 10 associated with the quality scores (since different read-haplotype pairs may have different scores and, therefore, require different powers of 10, which will be at different positions in the array). In lines 23 to 27, the haplotype characters are read from memory ($m256\_load$) and the first row of $M$ (line 26) and $I$ (line 27) is calculated. When the read and haplotype characters are compared (line 24), a mask is created ($m256\_cmpeq$) to select between the two possible values (line 25) that $P(r_i|h_j)$ can assume (see Equation \ref{eq:quality}). These two values are calculated on lines 20 and 21 to avoid recalculating $P(r_i|h_j)$ for each haplotype character, since they depend only on the quality score of the current read character. 

In lines 30 to 57, the remaining rows are calculated. At each iteration of the outer loop, the next read characters and quality scores are loaded from memory (lines 31 to 35). Next, the new values of $T_{MI}$ (line 39) and $T_{MM}$ (line 43), as well as the possible values for $P(r_i|h_j)$ (lines 40 and 41), are calculated. The first element of a $M$ row is set to 0 (line 44) due to the initialization in Equation \ref{eq:init} ($M_{1j}$ depends on $M_{0,j-1}$ and $I_{0,j-1}$, which are set to 0). The remaining elements are calculated in the inner loop (lines 46 to 54) according to Equation \ref{eq:dloop2}, with the last elements of $M$ and $I$ being updated in line 56. In line 59, the results of the last row of $M$ and $I$ are summed and stored in memory in line 61.

\subsection{GPU Cross-Thread Intrinsics} \label{sec:warp}

In Section \ref{sec:overview}, Endeavor's GPU approach was defined to process a read-haplotype pair at the warp-level by leveraging row-wise parallelism. Unlike CPUs, GPU threads cannot efficiently compute a complete row of $M$ and $I$ for long sequences, as keeping the results in private memory would lead to register spilling and lower performance. Therefore, each thread calculates $N$ elements of a row. However, note that the row-wise definition of PairHMM implies that not all elements of $M$ require the same amount of computation, which will hinder performance due to poor load balancing. For example, $M_{ij}$ needs $j-2$ elements to compute the sum derived in Equation \ref{eq:dloop}, $M_{i,j-1}$, requires $j-3$ elements, i.e., each subsequent $M_{ij}$ will require one less element. To overcome load balancing issues between different threads within the same warp, a warp shuffle strategy is proposed to ensure the correctness of $M$ and illustrated in Figure \ref{fig:general} with an example using 2 threads, each calculating 4 elements of $M$ (i.e., $N = 4$).

The first two elements ($M_{i,1}$ and $M_{i,2}$) do not compute the sum from Equation \ref{eq:dloop}, as it would start from $M_{i-1, -1}$ (which does not exist) and $M_{i-1, 0}$ (which is zero), respectively. $M_{i,3}$ needs $M_{i-1, 1}$ for the sum and the next element, $M_{i,4}$, needs $M_{i-1, 2} + T_{DD}M_{i-1, 1}$, where the second term is the sum from the previous calculated element of $M$ ($M_{i,3}$) scaled by $T_{DD}$. In general, $M_{i,k}$ scales the sum of the previous element, $M_{i,k-1}$, by $T_{DD}$ and adds another element to the sum. While thread 0, which calculates the first $N$ elements of $M$, can apply the consecutive additions and scaling operations and correctly calculate $M$, the same cannot be said for thread 1, which needs the sum from thread 0. In general, thread $T_{ID}$ needs the sums from threads 0 to $T_{ID} - 1$. The structure of the terms for each thread to calculate $M$ according to Equation \ref{eq:dloop} resembles the inclusive scan \cite{harris2007parallel}, apart from $T_{DD}$ factors. 

\begin{algorithm}[t]
\caption{Warp Inclusive Scan}
\begin{algorithmic}[1]
        \STATE \textbf{Input}: \textit{T} (value of thread), \textit{tid} (thread ID)
        \STATE \textbf{Output}: \textit{T} (result of prefix sum for each thread)\\

        \STATE
        
        \FOR{($offset = 1; \ offset < W; \ offset \ \times= \ 2$)}
            \STATE $\_aux = $\textbf{\_\_shfl\_inst}($-1, T, offset$);
            \STATE \textbf{if} $tid\%W \geq offset$: $T \ += \_aux \times T_{DD}^{offset*N}$;
        \ENDFOR
    
        \RETURN $T$

\end{algorithmic} 
\label{alg:warpprefix}
\end{algorithm}

The first row of $M$ ($M_{1,j}$) is easily calculated due to the algorithm's initialization (because $M_{0,j} = I_{0,j} = 0$ and $D_{0,j} = 2^{1020}/|H|$, $M_{1,j} = p_{1,j} \times T_{DM} \times 2^{1020}/k$). As each element is calculated, each thread accumulates the scaled sum ($V$). The following step makes use of intrinsic shuffle operations ($\_\_shfl\_inst()$) so that thread $T_{ID}$ receives $V$ from thread $T_{ID} - 1$ and stores the result in $T$ (for CUDA and HIP, this instruction is equivalent to $\_\_shfl\_up\_sync()$, and $\_\_shfl\_up()$, respectively). For thread 0, $T$ is set to 0. After setting $T$, warp shuffle operations are executed to perform an inclusive scan, as described in Algorithm \ref{alg:warpprefix}, in a warp or subwarp of size $W$. In Figure \ref{fig:general}, $W = 2$ and only one shuffle instruction is necessary, where thread $T_{ID}$ receives $T$ from thread $T_{ID} - 1$ and threads with $T_{ID} \geq 1$ add it to its own value of $T$ by scaling it to a factor of $T_{DD}^{N}$ (where $N = 4$ in the case of Figure \ref{fig:general}). In general, for a warp or subwarp of size $W$, $log_2(W)$ shuffle operations are necessary, scaled by powers of $T_{DD}^N$, as depicted in Algorithm \ref{alg:warpprefix}. After these operations, each thread has the correct sum for the next row of $M$. For example, for thread 1, $T$ is given by
\begin{equation}
    \begin{aligned}
    T & = \sum_{b=1}^{8}M_{1,b}T_{DD}^{8 - b}\\
    \end{aligned}
\end{equation}   
and the sums that thread 1 needs to calculate its elements ($M_{2,5}$ to $M_{2,8}$) are given by
\begin{equation}
    \begin{aligned}
    M_{2,9} & \rightarrow \sum_{b=1}^{7}M_{1,b}T_{DD}^{7 - b} = T_{DD}^{-1}(T - M_{1,8})\\
    M_{2,10} & \rightarrow \sum_{b=1}^{8}M_{1,b}T_{DD}^{8 - b} = T\\
    M_{2,11} & \rightarrow \sum_{b=1}^{9}M_{1,b}T_{DD}^{9 - b} = T_{DD}T + M_{1,9}\\
    M_{2,12} & \rightarrow \sum_{b=1}^{10}M_{1,b}T_{DD}^{10 - b} = T_{DD}^2T + T_{DD}M_{1,9} + M_{1,10}\\
    \end{aligned}
\end{equation}
which, as is shown, can be derived from $T$. For each thread, the first element of $M$ is done separately, as it requires values from other threads (in this example, thread 1 needs to retrieve $M_{1,4}$ from thread 0 to calculate $M_{2,5}$), which can be obtained through shuffle intrinsics. Afterwards, the remaining elements can be calculated in a loop, scaling $T$ with $T_{DD}$ while adding the necessary elements from the previous row and scaling $V$ with $T_{DD}$ while adding the necessary elements from the current row. As $T_{DD}$ is constant (0.1), all its powers are also constant and stored in constant memory.

Listing \ref{code:gpupairhmm} provides the developed GPU kernel code that implements Endeavor's row-wise PairHMM algorithm. The overall structure is similar to Listing \ref{code:cpupairhmm}, with $Mp$ and $Ip$ being defined as arrays of size $N$ for each thread, the addition of the warp inclusive scan (lines 21 and 45), which corresponds to the "Inclusive Scan" section in Figure \ref{fig:general}), the shuffle intrinsics that allow thread $T_{ID}$ to retrieve the last values for $M$ and $I$ from thread $T_{ID} - 1$ (lines 27 and 28), and the shuffle intrinsics to exchange $V$, which has the accumulated sum of $M$ elements of each thread (lines 19 and 43).

\begin{lstlisting} [float,style=cuda, caption={GPU kernel for warp-level PairHMM (FP64).}, label=code:gpupairhmm]
template <unsigned int N, unsigned int P> 
__global__ void fp64_subwarp(char* hap, char* _rs, char* _q, char* _i, char* _d, double* q64, double* finalsum)
{
  short subwarp = 32/P;
  short laneID = threadIdx.x%subwarp, warpID = threadIdx.x/subwarp; 
  int seqidx = P*blockIdx.x + warpID;
  double Mp[N] = {0}, Ip[N] = {0};
  char _hap[N] = {0};
  double M=0, I=0, y1=0, y2=0, T=0, V=0, d0=0, tmm=0;
  char rs, q, in, d1;

  rs = _rs[tidx], q = _q[tidx];
  for(unsigned int h = 0; h < N; h++){
    _hap[h] = hap[32*h + tidx];
    Mp[h] = (_hap[h] == rs) ? (1 - q64[q]) : q64[q]*0.33;
    Mp[h] *= 0.9*ldexp(1,1020)/N; 
    V = V*0.1 + Mprev[h];  
  }
  T = __shfl_inst(-1, V, 1);
  if(laneID == 0) T = 0;
  T = warp_inclusive_scan(T, laneID);
  for(int k = 1; k < N; k++){
    rs = _rs[32*k + tidx], q = _q[32*k + tidx]; 
    in = _i[32*k + tidx], d0 = q64[_d[32*(k-1) + tidx]]; 
    d1 = _d[32*k + tidx];
    tmm = 1 - q64[in] - q64[d1];
    V = __shfl_inst(-1, Mp[N-1], 1);
    y1 = __shfl_inst(-1, Ip[N-1], 1);
    y2 = (_hap[0] == rs) ?  (1 - q64[q]) : q64[q]*0.33;
    y2 *= (V*tmm + 0.9*(y1 + d0*10*(T - V)));
    if(laneID == 0) y2 = 0;
    M = y2, V = y2;
    I = Mp[0]*q64[in] + Ip[0]*0.1;
    for(int h = 1; h < N; h++){
      y2 = (_hap[h] == rs) ? (1 - q64[q]) : q64[q]*0.33;
      y2 *= (Mp[h-1]*tmm + 0.9*(Ip[h-1] + d0*T));
      T = T*0.1 + Mp[h-1];
      Mp[h-1] = M, Ip[h-1] = I;
      M = y2, I = Mp[h]*q64[in] + Ip[h]*0.1;
      V = V*0.1 + y2;    
    }       
    Mp[N-1] = M, Ip[N-1] = I; 
    T = __shfl_inst(-1, V, 1);
    if(laneID == 0) T = 0;
    T = warp_inclusive_scan(T, laneID);
  }
  y1 = 0;
  for(int h = 0; h < N; h++) y1 += Mp[h] + Ip[h]; 
  y1 = warp_inclusive_scan(y1, laneID);
  if(laneID == subwarp - 1) finalsum[seqidx] = y1;
}
\end{lstlisting}

The approach herein devised works for haplotypes of size up to $N \times W$ (where $N$ is the number of elements a single thread processes and $W$ the warp size). For bigger haplotypes, Endeavor can be scaled to use multiple warps. To achieve this, the last thread of each warp saves in shared memory its last element of $M$ and $I$ (to be used by the first thread of the next warp to calculate its first element of $M$), and the sum of all $M$ elements calculated by the warp, given by
\begin{equation}
    \begin{aligned}
    T & = \sum_{b=1}^{N \times W}M_{i,b}T_{DD}^{N \times W - b}\\
    \end{aligned}
\end{equation}  
as this result will be needed by the next warp to calculate its $M$ elements. To ensure correctness when reading from/writing to shared memory, thread synchronization is added.

\begin{figure*}[t] 
    \begin{subfigure}[t]{0.99\columnwidth}
    \includegraphics[width=\linewidth]{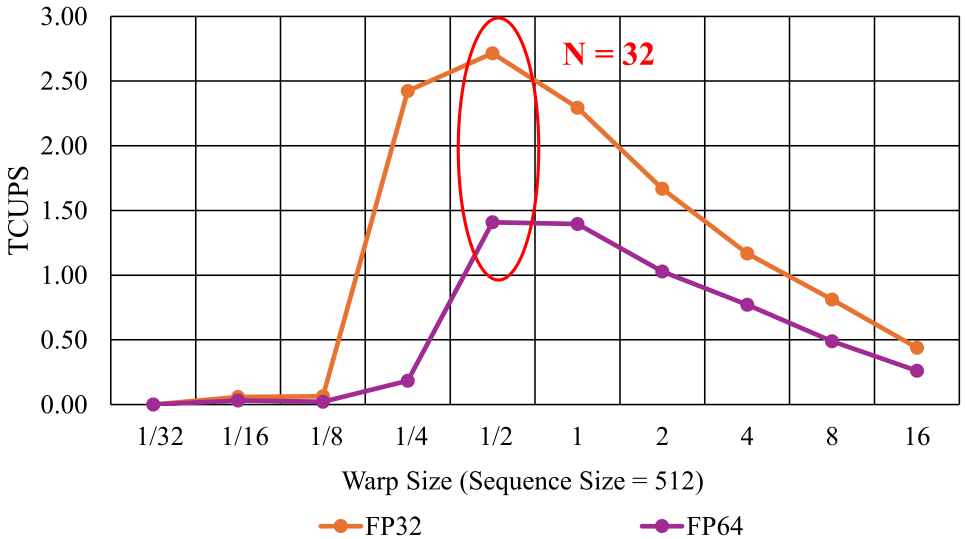}
    \caption{H100}
    \label{kernel:h100}
    \end{subfigure}
    \hfill
    \begin{subfigure}[t]{0.99\columnwidth}
    \includegraphics[width=\linewidth]{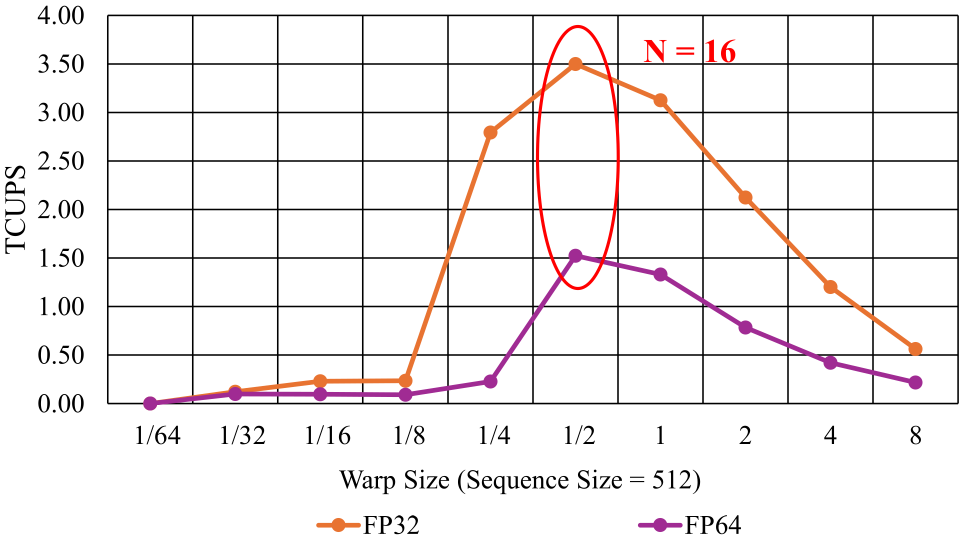}
    \caption{MI300A}
    \label{kernel:mi300a}
    \end{subfigure} 

    \caption{TCUPS evolution with the elements processed by each thread ($N$) in H100 (left) and MI300A (right) in FP32 and FP64.}
\end{figure*}

Setting $MT$ as the maximum number of threads in a block, this approach works for haplotypes up to $N \times MT$. To process even longer read-haplotype pairs, one can use distributed shared memory to exchange values between different thread blocks within a cluster. To achieve this, the last thread of each block saves in distributed shared memory its last element of $M$ and $I$ (to be used by the first thread of the next block to calculate its first element of $M$), and the sum of all $M$ elements calculated by the block, given by
\begin{equation}
    \begin{aligned}
    T & = \sum_{b=1}^{N \times MT}M_{i,b}T_{DD}^{N \times MT - b}\\
    \end{aligned}
\end{equation}
as this result will be needed by the next warp to calculate its $M$ elements. To ensure correctness when reading from/writing to shared memory, inter-block synchronization is added.

%% file: tex/Results.tex
In this section, details on the experimental evaluation of the Endeavor framework are provided. First, we focus on Endeavor's scalability and portability to answer the following questions: (i) how does the number of elements that each thread calculates ($N$) impact Endeavor's throughput on GPUs?, and (ii) how does the throughput scale with sequence length on different CPUs and GPUs? Second, Endeavor is tested on large-scale genomic datasets from different sequencing technologies, with varying sequence lengths and sequence pairs to be processed. In doing so, we answer the following questions: (i) how does Endeavor compare with a highly-optimized GATK HaplotypeCaller PairHMM CPU-based method?, and (ii) how does Endeavor compare with state-of-the-art GPU-based methods? 

\subsection{Experimental Setup} \label{sec:expsetup}

The reported experiments have been performed on 5 GPUs from different major vendors, i.e., NVIDIA V100 (300 W), A100 (300 W), H100 (700 W), and RTX 6000 PRO (600 W) GPUs, using the CUDA 12.8 toolkit, and AMD MI300A (750 W), using ROCM 6.3.4, as well as 2 CPU systems, with AMD EPYC Zen 4 (which is included in MI300A) and Intel Xeon Gold 6438 architectures, both with 96 cores and 192 available threads.  

To fully examine the capabilities of the proposed approach in datasets generated with short-read and long-read sequencing technologies, datasets from the GIAB \cite{xiao2014giab} consortium are used, namely the NA12878 \cite{NIHIllumina, NIHIon, NIHPacBio, NIHSolid}, NA24695 \cite{NIHONT}, NA24631 \cite{NIHBGI}, and NA24149 \cite{NIHChr} datasets. For the GIAB datasets, Endeavor is tested on all GPUs and compared with gpuPairHMM \cite{schmidt2024gpupairhmm} (as it is the current best PairHMM implementation for GPU) and GATK HaplotypeCaller, which uses the GKL \cite{foley2017accelerate} implementation for PairHMM and is capable of processing long-read datasets, on the Intel Xeon Gold 6438 CPU (with AVX-512 instructions) using all available threads (192). The 10s dataset \cite{carneiro2013accelerating} is also used for benchmarking and comparison with other approaches in the literature.

The evaluation of Endeavor and comparison to other state-of-the-art solutions focuses on throughput, measured as Tera Cell Updates Per Second (TCUPS) and calculated as
\begin{equation*}
    TCUPS = \frac{\sum_i^s (rlen_i \times hlen_i)}{t \times 10^{12}},
\end{equation*}
where $t$ is the runtime in seconds, $s$ is the total number of read-haplotype pairs to process, $rlen_i$ and $hlen_i$ are the length of the $i$-th read and haplotype, respectively. For benchmarking, varying sequence lengths are not considered (i.e., all read-haplotype pairs are the same length). For the GPU approach, for fair comparison with the state-of-the-art \cite{schmidt2024gpupairhmm}, in Section \ref{sec:scale}, only the kernel runtime is considered, excluding data transfer overheads, while in Section \ref{sec:large} the data transfer time is also considered in the execution times.

\subsection{GPU Kernel Parameters Analysis} \label{sec:kernel}

In this section, an analysis on the impact of $N$ (the number of elements of $M$ and $I$ processed per thread) in GPU performance is performed. To this end, experiments on H100 and MI300A are conducted, fixing the read-haplotype size to 512 (this is in the middle of the short-read range, which typically spans 250-800 basepairs \cite{hu2021next}) and varying the value of $N$, using the warp size as a reference. Among the tested NVIDIA GPUs, H100 is the best choice to analyze both precisions (while RTX 6000 PRO has higher FP32 peak throughput, the FP64 peak throughput is 1/64 of the single precision, instead of 1/2 as is the case of H100). Figures \ref{kernel:h100} and \ref{kernel:mi300a} show the variations in TCUPS for H100 and MI300A, respectively, on FP64 and FP32, according to $N$. H100 has a warp size of 32 (with $1/32$ representing a single thread), while MI300A has a warp size of 64 (with $1/64$ representing a single thread).

If $N$ is too small, the number of threads per block has to be increased, with each thread performing little work, resulting in decreased throughput. On the other hand, if $N$ is too big (in the limit, if a single thread calculates all values of $M$ and $I$ for a single read-haplotype pair), the register pressure results in spilling to the global memory, which, due to its high latency, reduces the kernel's throughput. Therefore, one must achieve a balance between providing enough elements for threads to process and ensuring that register spilling does not occur. For H100, the optimal value is $N = 32$ (i.e., for the considered length of 512, 16 threads, which is 1/2 of a warp, calculate 32 elements of $M$ and $I$ rows), followed by $N = 64$, while MI300A achieves the best TCUPS for $N = 16$. The results show that the optimal value of $N$ is hardware dependent and can be tuned to achieve maximum performance on different GPU architectures. For this work, V100, A100, H100, and RTX 6000 PRO use $N = 32$ for sequence lengths up to 1024 and $N = 64$ for longer sequences, while MI300A uses $N = 16$ for all sequence lengths.

\subsection{Scalability Analysis} \label{sec:scale}

To evaluate Endeavor's scalability, synthetic datasets with sequence lengths ranging from 32 to 131072 basepairs are generated, with the read and haplotype having the same length. In doing so, we focus on answering the following questions: (i) how does Endeavor's performance scale on CPUs and compares to the state-of-the-art libraries?, and (ii) how does Endeavor perform on GPUs from different vendors and compares to the state-of-the-art approaches?

\subsubsection{CPU Analysis} 
\begin{figure}[t]
\centering
\includegraphics[width=\linewidth]{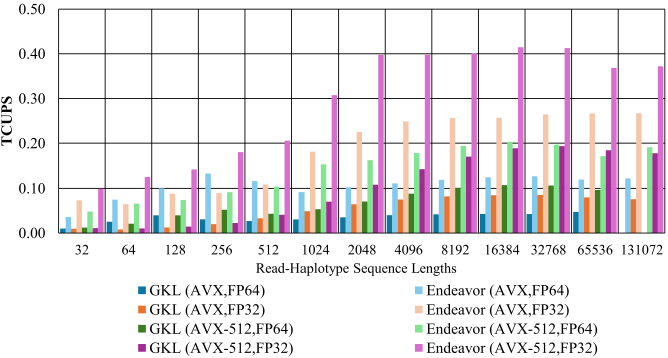}
\caption{TCUPS evolution in Intel Xeon Gold 6438 for Endeavor and GKL.} 
\label{fig:intelcpu}
\end{figure}

\begin{figure}[t]
\centering
\includegraphics[width=\linewidth]{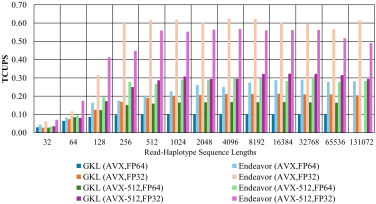}
\caption{TCUPS evolution in AMD EPYC Zen 4 for Endeavor and GKL.} 
\label{fig:amdcpu}
\end{figure}

Figures \ref{fig:intelcpu} and \ref{fig:amdcpu} display the TCUPS evolution in the Intel and AMD CPUs, respectively, as the input read-haplotype length increases. Eight bars are displayed, four for Endeavor combining FP32/FP64 precisions and AVX/AVX-512 SIMD widths, and four for the CPU state-of-the-art approach, GKL \cite{foley2017accelerate}, for the same precisions and SIMD widths. The results demonstrate that Endeavor outperforms GKL for all tested combinations of SIMD width and floating-point precisions, with the AMD EPYC Zen 4 achieving a peak throughput of 0.625 TCUPS using AVX with FP32 and the Intel Xeon Gold 6438 achieving the best results for AVX-512 and FP32, with 0.415 TCUPS. In contrast, GKL only achieves a peak throughput of 0.323 TCUPS and 0.194 TCUPS, respectively, using AVX-512 with FP32 in both cases. Therefore, Endeavor achieves a 1.93x and 2.14x improvement in PairHMM throughput. Note that Zen 4 does not achieve optimal throughput using AVX-512 as the operations run on a 256-bit data path that requires two passes to process 512 bits \cite{bhargava2024amd}, which results in lower performance. 

As the sequence length increases, the throughput for both approaches tends to increase and hit a peak for long reads. As an example, for a sequence length of 16384, Endeavor achieves between 1.69x and 2.83x speedups for the AMD CPU, while Intel CPU achieves speedups between 1.89x and 3.03x. For both devices, the lower speedups occur for AVX-512 with FP64 precision, while the higher speedups occur for AVX with FP32 precision. Finally, note that the proposed approach is scalable to long reads for any precision and any SIMD width, while GKL fails to run in FP64 for the last tested sequence length (131072) in any SIMD width.

\subsubsection{GPU Analysis} 

%



\begin{figure}[t]
\centering
\includegraphics[width=\linewidth]{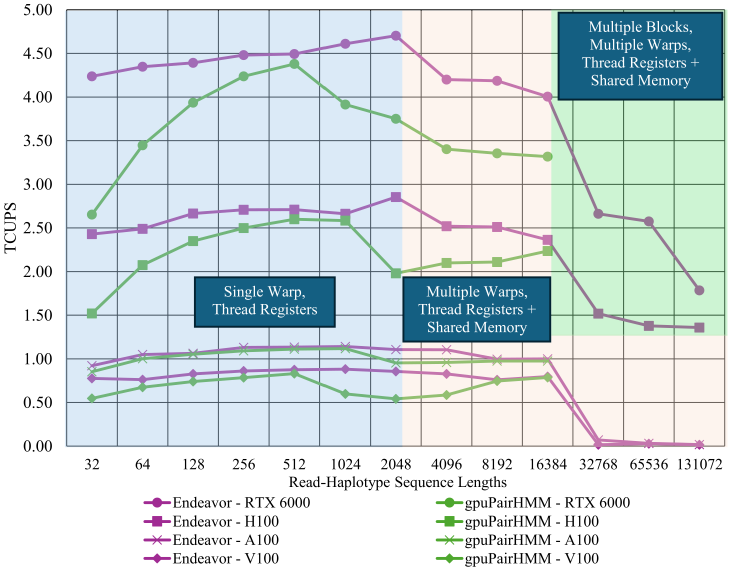}
\caption{TCUPS evolution in V100, A100, H100, and RTX 6000 for Endeavor and gpuPairHMM.} 
\label{fig:scalenvidia}
\end{figure}

Figure \ref{fig:scalenvidia} displays the TCUPS evolution in the V100, A100, H100, RTX 6000 GPUs in FP32 as the input read-haplotype length increases. Eight curves are displayed, four for Endeavor (purple) and four for the GPU state-of-the-art approach, gpuPairHMM \cite{schmidt2024gpupairhmm}. MI300A is not included, as gpuPairHMM only runs in NVIDIA GPUs. As mentioned in Section \ref{sec:kernel}, Endeavor uses $N = 32$ for FP32 in NVIDIA GPUs for sequence lengths up to 1024 and $N = 64$ for longer read-haplotype pairs. Given the warp size of 32, a single warp can process read-haplotype pairs of length up to 2048 with FP32 (blue area). For long read-haplotype pairs, processing is done at the level of multiple warps, leveraging shared memory to store necessary values across warps (orange area) and at the level of multiple thread blocks for very long reads (green area), using thread block clusters, which are only available in H100 and RTX 6000. Therefore, for sequences longer than 16384, V100 and A100 still rely on shared memory and multiple warps. As gpuPairHMM's benchmark only runs for sequence lengths up to 16384, its curve is not represented for longer read-haplotype pairs.

For older GPU architectures (V100 and A100), Endeavor is on par with gpuPairHMM, with more significant improvements in throughput for sequence lengths between 1024 and 4096, where Endeavor achieves at least 1.4x speedup on V100 and 1.16x speedup on A100. For sequence lengths higher than 16384, the throughput is significantly lower due to increased register pressure and spilling to global memory. For more recent GPU architectures (H100 and RTX 6000), Endeavor achieves a peak throughput of 2.85 TCUPS and 4.70 TCUPS, while gpuPairHMM achieves 2.6 TCUPS and 4.38 TCUPS, which represents a 9.6\% and 7.3\% improvements, respectively. For these two GPUs, speedups are more significant in the 32-128 range, with 1.6x speedups for sequence lengths of 32 basepairs, and in the 2048-8192 range, where both devices achieve at least 1.2x speedups. Finally, Endeavor achieves good performance for sequence lengths up to 131072 basepairs, which is one order of magnitude higher than the largest sequence length that the GPU state-of-the-art approaches can process.

\begin{figure}[t]
\centering
\includegraphics[width=\linewidth]{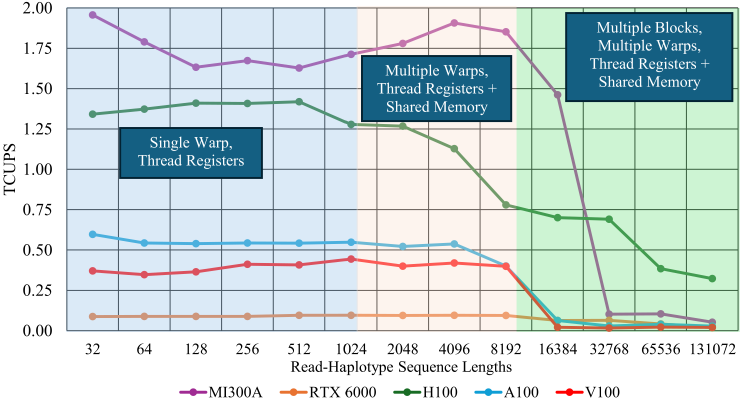}
\caption{TCUPS evolution with sequence length in Endeavor (FP64).} 
\label{fig:scalefp64}
\end{figure}

\begin{figure}[t]
\centering
\includegraphics[width=\linewidth]{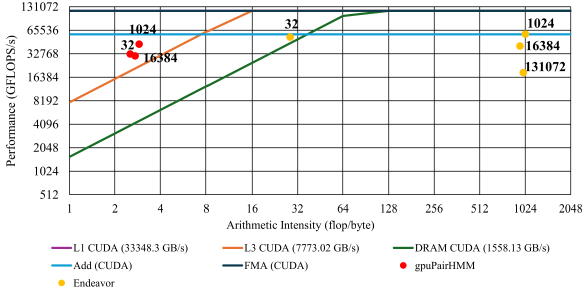}
\caption{CARM Roofline for Endeavor and gpuPairHMM.} 
\label{fig:carmtool}
\end{figure}

Figure \ref{fig:scalefp64} provides Endeavor's FP64 results on all tested GPUs. Note that gpuPairHMM is not represented in this graph as it only supports single precision. The results show that MI300A achieves the best peak throughput in double precision (1.95 TCUPS), followed by H100 (1.42 TCUPS), A100 (0,6 TCUPS), V100 (0.44 TCUPS) and RTX 6000 PRO (0.01 TCUPS). As mentioned in Section \ref{sec:kernel}, the FP64 throughput for RTX 6000 PRO is 1/64 of the peak FP32 throughput, which explains the lower TCUPS when compared to other GPU devices. MI300A achieves the best results on sequence lengths up to 16384. For longer sequences, the throughput is significantly lower due to increased register pressure, similarly to V100 and A100. For sequence lengths between 32768 and 131072, H100 achieves the best throughput, as it leverages thread block clusters to maintain high performance.

In FP32, considering the peak throughputs and TDP of each device, RTX 6000 PRO is the most energy-efficient, providing 0.0078 TCUPS/W, followed by H100 (0.004 TCUPS/W), A100 (0.0038 TCUPS/W), and V100 (0.0029 TCUPS/W). In FP64, MI300A is the most energy-efficient (0.0026 TCUPS/W), followed by H100/A100 (0.002 TCUPS/W), V100 (0.0015 TCUPS/W) and RTX 6000 PRO (0.00016 TCUPS/W). Therefore, Endeavor's portability allows to choose an optimal device considering different metrics (energy efficiency, targeted sequence length, throughput, or precision).

Finally, to further demonstrate Endeavor's efficiency in using the GPU's resources, Figure \ref{fig:carmtool} displays the RTX 6000's roofline (as it is the evaluated device with higher FP32 throughput), generated with CARMTool \cite{morgado2024carm}, to evaluate Endeavor (for sequence lengths of 32, 2048, 16384, and 131072) and gpuPairHMM (for sequence lengths of 32, 1024, and 16384), with varying sequence lengths to evaluate the performance of both methods as the input size increases. The roofline shows that gpuPairHMM is closer to the memory-bound region (the main limitations come from the memory subsystems, which hinders possible performance improvements), while Endeavor, as a redefined PairHMM approach, is compute-bound (the main bottleneck is the amount of compute resources in the hardware).

\begin{table}[t]
\setlength{\tabcolsep}{2pt}
\renewcommand{\arraystretch}{1}
\centering
{\fontsize{8pt}{10pt}\selectfont
\caption{GIAB Dataset Description.}
\label{tab:shortread}
\begin{tabular}{llllll}
\hline
Sample & Technology & \#Pairs & Min & Median & Max\\
\hline
NA12878 \cite{NIHSolid} & SoLiD & 113960218 & 10 & 50 & 481\\
NA24631 \cite{NIHBGI} & BGISEQ-500 (chr 22) & 158287379 & 10 & 149 & 498\\
NA12878 \cite{NIHIllumina} & Illumina & 1442661880 & 10 & 145 & 897\\
NA12878 \cite{NIHIon} & Ion Torrent & 1672115436 & 10 & 208 & 499\\
NA24149 \cite{NIHChr} & Chromium (chr 22) & 131420448 & 10 & 151 & 651\\
NA12878 \cite{NIHPacBio} & PacBio & 866041959 & 10 & 215 & 3396\\
NA24695 \cite{NIHONT} & ONT (chr 22) & 284136274 & 10 & 299 & 12121\\
\hline
\end{tabular}
}
\end{table}

\subsection{Evaluation on Real Datasets} \label{sec:large}

In this section, experiments on GIAB datasets from seven different sequencing technologies are performed to evaluate Endeavor's flexibility and performance on short-read datasets (sequence lengths in the range of 250-800) and long-read datasets (sequence lengths $> 10^4$). Table \ref{tab:shortread} provides a description of the GIAB short-read and long-read datasets that are tested. For short-reads, four different datasets are considered, generated by Illumina, SoLiD, BGI-SEQ500, and Ion Torrent, with varying number of sequence pairs to process. For long-reads, three different datasets are evaluated, generated by Chromium Long Ranger, PacBio, and ONT sequencing technologies, with the last one exhibiting the sequences with longest length (12121 basepairs). The BGI-SEQ500, ONT and Chromium datasets refer only to Chromosome 22, while the remaining datasets refer to the complete human genome. 

The analysis of these datasets is done in three parts. First, Endeavor's CPU approach is compared to GATK's PairHMM AVX-512 implementation, which runs on the Intel Xeon Gold 6438 CPU, using 192 threads, in the seven GIAB datasets. Second, Endeavor's GPU approach is compared to gpuPairHMM in all GIAB datasets and datasets from the original paper \cite{schmidt2024gpupairhmm}. Finally, the 10s dataset \cite{carneiro2013accelerating}, a typical benchmark used in the literature to test novel approaches for PairHMM, is also evaluated with Endeavor and compared to gpuPairHMM and GKL, as well as other well-known PairHMM implementations in the literature (based on CDP \cite{liu2023genomics} and on inter- and intra-task parallelization \cite{ren2018efficient}).

\subsubsection{Comparison with GATK HaplotypeCaller}

\begin{figure}[t]
\centering
\includegraphics[width=\linewidth]{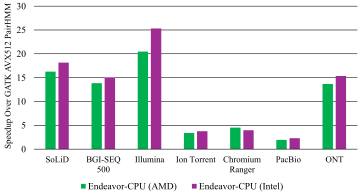}
\caption{Speedup of Endeavor-CPU over GATK AVX-512 (higher is better)}
\label{fig:gatk512pair}
\end{figure}

Figure \ref{fig:gatk512pair} provides the speedups achieved over GATK HaplotypeCaller's AVX-512 implementations on all datasets across the tested CPU devices. For a fair comparison, Endeavor runs AVX-512 with single precision and the same number of threads as GATK HaplotypeCaller (192). On the AMD CPU, Endeavor's speedups range from 1.94x to 20.48x, while on the Intel CPU, the speedups are between 2.29x and 25.32x, with the lowest results for the PacBio dataset and the highest results for the Illumina dataset. The average speedups for both devices are 10.59x (AMD CPU) and 11.98x (Intel CPU).

\subsubsection{Comparison with GPU-based methods} \label{sec:sotacomp}
\begin{figure}[t]
\centering
\includegraphics[width=\linewidth]{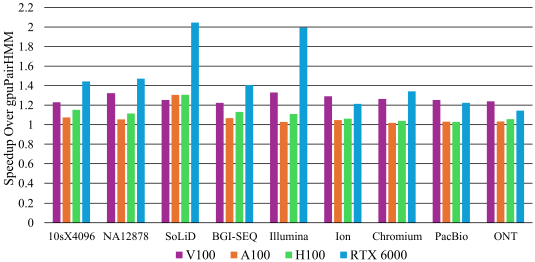}
\caption{Speedup of Endeavor-GPU over gpuPairHMM (higher is better).}
\label{fig:sotacomp}
\end{figure}
Figure \ref{fig:sotacomp} provides the speedups achieved over gpuPairHMM with the Endeavor framework on the GIAB, in addition to datasets considered in the gpuPairHMM \cite{schmidt2024gpupairhmm} paper: a subset of the NA12878 dataset from the 1000 Genomes project \cite{10002015global} and the 10s dataset \cite{carneiro2013accelerating} replicated 4096x. The results show that Endeavor provides slight improvements over gpuPairHMM in A100 and H100 on most of the datasets (with the exception of the SoLiD dataset, where Endeavor achieves the highest speedup, 1.3x). For V100, Endeavor achieves at least 1.22x speedups, while for the newest GPU architecture (RTX 6000 PRO), Endeavor consistently outperforms gpuPairHMM, with speedups between 1.14x (ONT) and 2.05x (SoLiD), and an average 1.47x speedup.

Note that, as Endeavor's results match the results from gpuPairHMM \cite{schmidt2024gpupairhmm} and GATK HaplotypeCaller \cite{depristo2011framework}, substituting the latter's CPU-based implementation with Endeavor's GPU approach would result in significant speedups with no loss in precision. When compared to GATK HaplotypeCaller, the achieved speedups range between 15.66x-229.96x (V100), 20.88x-305.04x (A100), 48.74x-674.75x (H100), 79.58x-1115.91x (RTX 6000), and 48.82x-758.35x (MI300A), with the lowest speedups on the Ion Torrent dataset and the highest on the SoLiD dataset.

\subsubsection{Comparison using the 10s dataset}

In this section, the 10s dataset \cite{carneiro2013accelerating} is evaluated on CPUs and GPUs. Although 10s is not representative of a real large-scale dataset, it is a benchmark that is widely used in the literature \cite{branchini2021methodology, ren2018efficient, schmidt2024gpupairhmm, sampietro2018fpga, banerjee2017accelerating, huang2017hardware} to test novel approaches for PairHMM. To examine Endeavor's performance on this dataset in CPUs, GKL is employed for comparison, while for GPUs, gpuPairHMM, as well as a benchmark for NVIDIA GPUs known as Genomics-GPU \cite{liu2023genomics}, are evaluated. Genomics-GPU is a software suite that includes a CDP implementation of PairHMM, as well as the implementation presented in \cite{ren2018efficient}, which exploits inter-task and intra-task parallelism to increase the algorithm's throughput. Table \ref{tab:10s} provides the execution times for the 10s dataset in all the PairHMM codes, including Endeavor.
\begin{table}[t]
\setlength{\tabcolsep}{2pt}
\renewcommand{\arraystretch}{1}
\centering
{\fontsize{8pt}{10pt}\selectfont
\caption{Execution Times for 10s Dataset (in milliseconds).}
\label{tab:10s}
\begin{tabular}{llllll}
\hline
Method & \multicolumn{3}{c}{Xeon Gold 6438} & \multicolumn{2}{c}{EPYC Zen 4}\\
\hline
GKL \cite{foley2017accelerate} & \multicolumn{3}{c}{15.39} & \multicolumn{2}{c}{23.95}\\
\textbf{Endeavor-CPU} & \multicolumn{3}{c}{\textbf{11.81}} & \multicolumn{2}{c}{\textbf{17.24}}\\
\hline
 Method & V100 & A100 & H100 & RTX 6000 & MI300A\\
\hline
Inter-task \cite{ren2018efficient} (Tile = 1) & 25.4 & 19.4 & 4.60 & 16.5 & - \\ 
Inter-task \cite{ren2018efficient} (Tile = 2) & 14.9 & 11.8 & 8.81 & 11.2 & - \\ 
Inter-task \cite{ren2018efficient} (Tile = 4) & 8.59 & 6.75 & 5.20 & 6.26 & - \\ 
Inter-task \cite{ren2018efficient} (Tile = 6) & 7.51 & 5.58 & 4.94 & 4.98 & - \\ 
Inter-task \cite{ren2018efficient} (Tile = 8) & 7.35 & 6.36 & 3.65 & 4.33 & - \\ 
Intra-task \cite{ren2018efficient} (Tile = 1) & 4.00 & 3.23 & 2.32 & 2.62 & - \\ 
Intra-task \cite{ren2018efficient} (Tile = 2) & 4.54 & 3.81 & 2.83 & 2.80 & - \\ 
Intra-task \cite{ren2018efficient} (Warp) & \textbf{1.58} & \textbf{1.25} & \textbf{0.91} & \textbf{0.76} & - \\ 
CDP \cite{liu2023genomics} & 13.2 & 14.6 & 7.59 & 5.94 & - \\ 
gpuPairHMM \cite{schmidt2024gpupairhmm} & 8.50 & 7.41 & 1.77 & 1.88 & - \\ 
\textbf{Endeavor-GPU} & \textbf{0.37} & \textbf{0.26} & \textbf{0.20} & \textbf{0.15} & \textbf{0.26}\\
\hline
\end{tabular}
}
\end{table}

Because the 10s dataset is small (3550 sequence pairs, with the longest sequence having 263 basepairs), it is possible to process all sequence pairs simultaneously. By doing so, Endeavor achieves 11.81 and 22.24 milliseconds on Intel Xeon Gold 6438 and AMD Zen 4 CPUs, respectively, outperforming GKL by 1.3x and 1.39x. On GPUs, Endeavor achieves 0.37, 0.26, 0.20, and 0.15 milliseconds in V100, A100, H100, and RTX 6000, which leads to a speedup of 4.27x, 4.81x, 4.55x, and 5.06x, respectively, over the second fastest solutions presented in the table (which are in bold). While Genomics-GPU and gpuPairHMM only run on NVIDIA GPUs, Endeavor can also be tested on MI300A, which takes 0,26 milliseconds to process 10s, showcasing Endeavor's portability across different devices.

%% file: tex/Conclusion.tex
Variant calling is a fundamental problem in bioinformatics with a significant impact on precision medicine, pharmacogenomics, and evolutionary biology. Given its importance, the development of efficient solutions to keep up with its computational demand is paramount. To that end, in this work, Endeavor, a novel parallelization strategy, was presented to leverage row-level parallelization of the PairHMM algorithm and achieve better usage of CPU's and GPU's hardware resources. The results on Intel and AMD CPUs showed up to 2.14x improvement in peak PairHMM throughput, as well as up to 25.32x speedup when processing genome-scale datasets. On NVIDIA and AMD GPUs, the proposed solution achieves up to 2.05x speedups over state-of-the-art GPU-based methods on the most recent GPU architectures. Finally, the proposed framework can run on short- and long-read datasets, demonstrating its capability to run on datasets from any sequencing technology and any sequence length on CPU and GPU devices.

%% file: tex/Acks.tex
This work was supported by national funds through Fundação para a Ciência e a Tecnologia, I.P. (FCT) under projects UID/50021/2025 (DOI: \url{https://doi.org/10.54499/UID/50021/2025}) and UID/PRR/50021/\allowbreak2025 (DOI: \url{https://doi.org/10.54499/UID/PRR/50021/2025}), LISBOA\allowbreak2030-FEDER-00869000 (2023.18110.ICDT, VERSACOMP, DOI: \url{https://doi.org/10.54499/2023.18110.ICDT}), and the UI/BD/154603/2022 research grant. We also acknowledge the European Union HE Research and Innovation programme under grant agreement No 10109\allowbreak2877 (SYCLOPS), and the EuroHPC Joint Undertaking for awarding us access to MareNostrum5 at BSC, Spain.